\documentclass[fleqn,twocolumn]{olplainarticle}
\usepackage{hyperref}       
\graphicspath{{./Figures}}
\usepackage{float}
\usepackage{xcolor}
\usepackage{ragged2e}
\usepackage{fixltx2e}
\usepackage{siunitx}
\usepackage{tabularx}
\usepackage{listings}
\usepackage{ragged2e}
\usepackage{hyperref} 
\usepackage{indentfirst}

\title{CHESSIoT support of event-based modeling for the Internet of Things applications}

\author{Felicien Ihirwe}
\vspace{-.6cm}
\affil{Address\\\
Email: felicien.ihirwe@intecs.it\\\
Innovation and technology Service (ITS) Lab\\\ Intecs Solutions Spa\\\
Pisa, Italy}

\begin{abstract}
Internet of Things systems design and development suffers from heterogeneity in different aspects. The component behaviors also change due to events being internal or external and the system needs to take action subsequently. In this paper, we demo the event-based modeling capabilities of the CHESSIoT tool on a smart parking application. Different components was been decomposed with functional and behavioral specifications following the component-based approach. In the end, we have given a brief description of the future work. 
\end{abstract}
\vspace{-.6cm}
\begin{document}

\flushbottom
\maketitle
\thispagestyle{empty}

\section{Background}
Finding a vacant parking slot in a big parking lot can be complex and time-consuming especially during peak periods. More complexes offer indoor parking for their clients but few of them provide an easy way to identify a vacant post for parking. Hence the clients are forced to drive through the parking lot looking for a vacant space in which they could end up not finding any at all.  To address this issue, this document proposes a sensor-based smart parking system.

A typical indoor smart parking system is composed of several nodes deployed into different parking spaces with a consolidated central unit. The deployed nodes collect the parking information to be used in direct decision-making or for further analysis purposes. The two main technologies involved in the modern indoor parking system car detection and identification are visual and sensor-based. The more recent technologies are adopting computer vision technologies but it is very expensive.  Depending on the technology used, the deployable nodes can be installed at a parking lot or at each parking slot in a parking lot. For this exercise, a sensor-based technology is used. The next figure \ref{fig:bigPicture} gives an show a \textbf{parkeasy} \footnote{\url{https://www.exportersindia.com/xaimen-parkeasy/}} sensor-based smart parking system.

\begin{figure}[h]
\centering
\includegraphics[width=\linewidth]{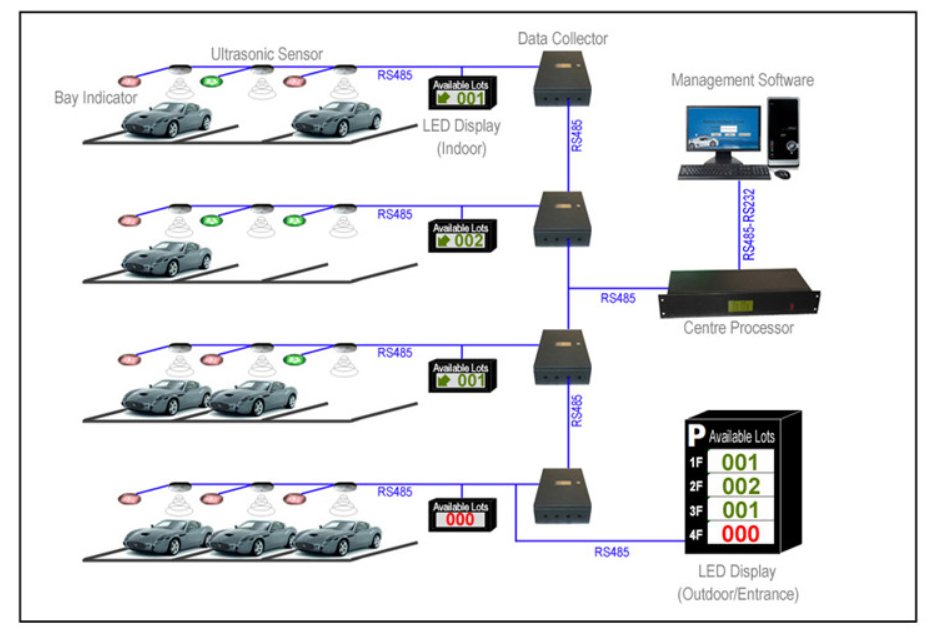}
\vspace{-.6cm}
\caption{Multi-level indoor parking system}
\label{fig:bigPicture}
\vspace{-.7cm}
\end{figure}

From the figure \ref{fig:bigPicture}, the system is composed of many different sensor nodes deployed on top of parking slots. Each parking slot is composed of an ultrasonic sensor fixed on the ceiling or on the floor of each parking slot. Ultrasonic sensors work based on echolocation technology in which a sensor transmits a sound wave to a given space and waits for an echo wave back. The time between the sent pulse and the returned echo is used to calculate distance and can give an idea if a slot is vacant or not. In a vacant space, the time between transmitted sound and reflection is longer than in an occupied space, hence the sensor can detect when space is occupied or not. The object’s distance is recalculated taking into account the speed of a sound wave. A node also has two separate LED indicators (RED and GREEN) for indicating the occupancy of the slot, so, in case GREEN is ON  means the parking slot is vacant while RED being ON means the slot is occupied.

The modeling of a system’s node was done using the CHESSIoT environment, a CHESS extension to cover the modeling and analysis of IoT systems. The full description of CHESSIoT can be found in our previous work \citep{CHESSIoT}. CHESS is a cross-domain, model-driven, component-based, and open-source tool for the development of high-integrity systems \citep{CHESSComplex}. For the sake of simplicity, this paper showcase the capability of CHESSIoT event-based behavioral modeling on a single slot. 
\begin{figure*}[h!]
\centering
\includegraphics[width=\linewidth]{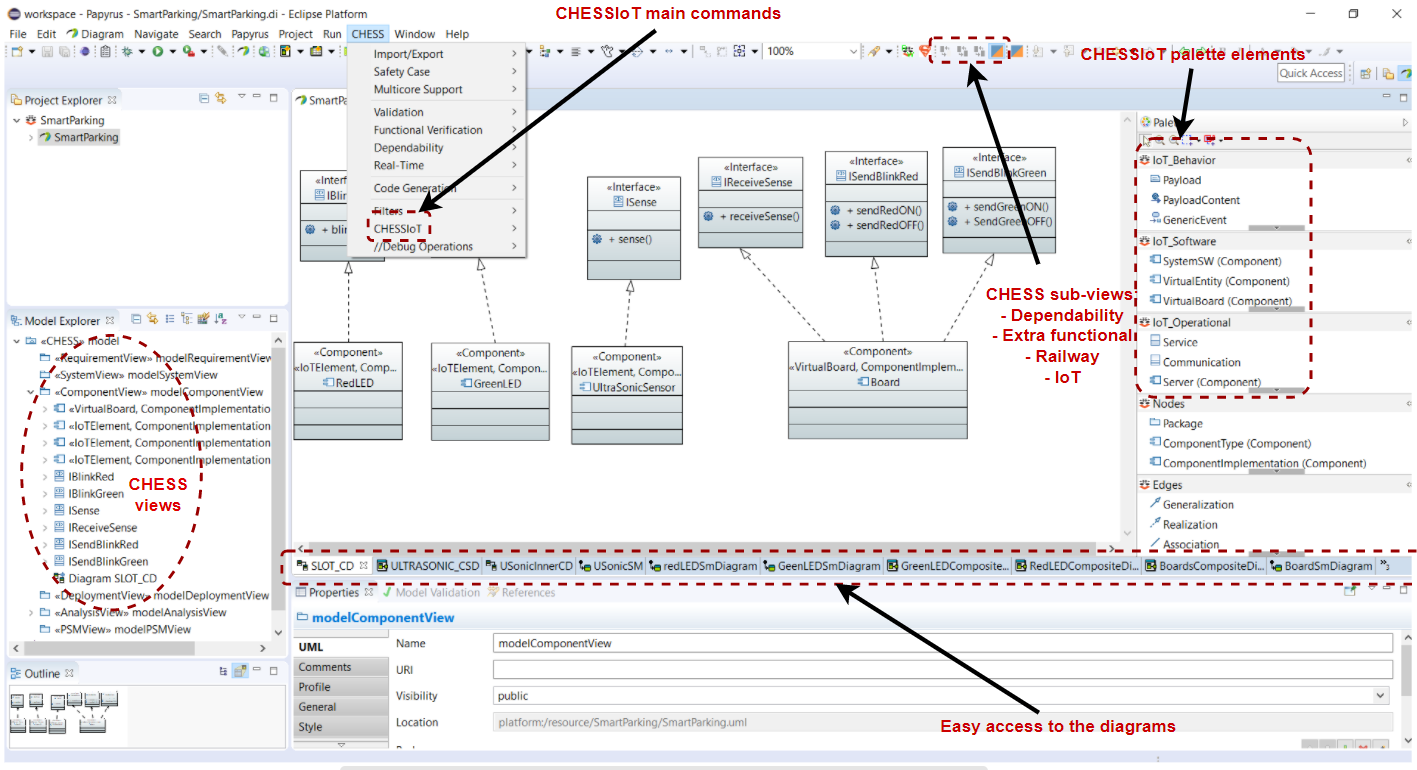}
\caption{CHESS editor structure}
\label{fig:chesseditor}
\vspace{-.3cm}
\end{figure*}

\section{CHESSIoT modeling editor}
CHESSIoT adds different IoT-related modeling functionalities on top of CHESS such as "IoT sub-view", "palletes" and several CHESSIoT specific commands to cover IoT modeling activities in CHESS. The following figure depicts the CHESS editor with the CHESSIoT extension.

\section{Software modeling}
The modeling of software components in CHESSIoT follows the component-based modeling approach enforced by CHESS. In this manner, the software main components are decomposed separately with their internal functionalities and behaviors and later be linked together to fulfill a certain task. The following figure depicts the physical structure of a node.

\begin{figure}[h]
\centering
\includegraphics[width=\linewidth]{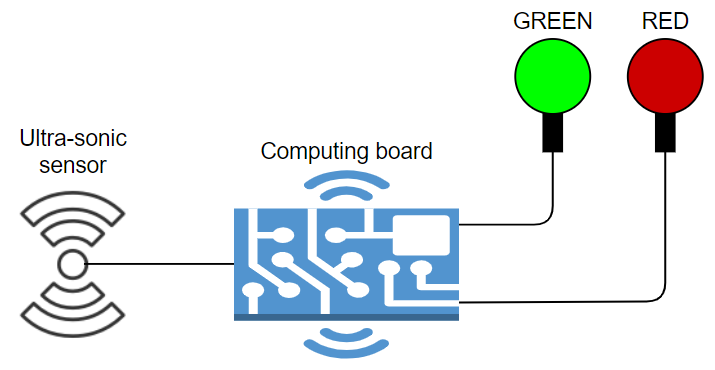}
\caption{ The physical structure of a node}
\label{fig:nodeStructure}
\vspace{-.6cm}
\end{figure}

\subsection{Components definition}
\begin{figure}[h]
\centering
\includegraphics[width=\linewidth]{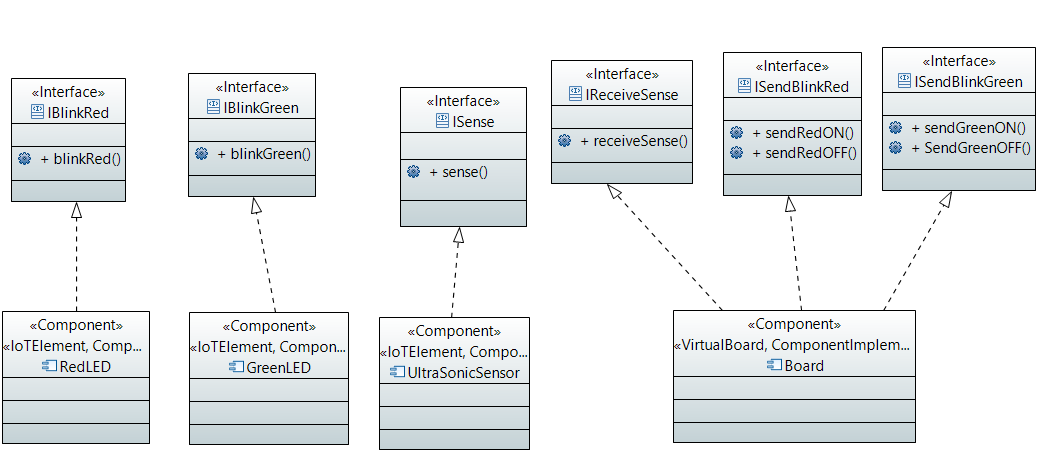}
\caption{A node software components and interfaces}
\label{fig:components}
\end{figure}

Modeling the software components is done in CHESS’s \textit{component view}.  In CHESSIoT, the software components such as \textit{IoTElement, VirtualBoard, VirtualEntity} are the main modeling elements. They are used to encapsulate the system's main part structure, operations, and behaviours. The process starts by defining the system's main components, interfaces, and operations. This is done by using UML class diagrams. Figure \ref{fig:components}. shows different main components involved in a node. From the figure \ref{fig:components}, and as mentioned in the previous section, the following components are defined as the main components for a node.

\textbf{RedLED} component is of  \textless\textless{}\textit{IoTElement}\textgreater\textgreater type and it represents an red LED indicator. This element implements the \textit{"IBlinkRed"} interface which imposes the \textit{“blinkRed()”} operation.

\textbf{GreenLED}: this is of type \textless\textless{}\textit{IoTElement}\textgreater\textgreater and it represents an green indicator. The same as the RedLED, GreenLED implements the \textit{“IBlinkGreen”} interface which imposes the \textit{“blinkGreen()”} operation.
(The above two elements will act as consumers, so they won’t be initiating any event by themselves)

\textbf{UltrasonicSensor}: this is of type \textless\textless{}\textit{IoTElement}\textgreater\textgreater and it represents an sensor. This element implements the \textit{“ISense”} interface which imposes the \textit{“sense()”} operation.

Board represents a computing device such as Arduino, Raspberry Pi,... It is of type \textless\textless{}\textit{Board}\textgreater\textgreater and realizes 3 different interfaces to interact with the LEDs and a sensor. Each interface exposes different operations depending on the action required.

\subsection{Modeling components internal structure}
To abide by the component-based approach each component’s internal structure needs to be decomposed separately by specifying their corresponding internal sub-components (if any) and ports. Ports are used to support the communication between two or more components exposing or requiring the interfaces from other components. In CHESSIoT, the component's messages are passed through the port using the required or provided interface operations. The component also has an associated initial state. The modeling of behavior information is modeled using a UML state machine and will be discussed later in this section. The modeling of internal structure is performed in the component view using their custom composite structure diagram. 

\subsubsection{Red/Green LED}
\begin{figure}[h]
\centering
\includegraphics[width=0.9\linewidth]{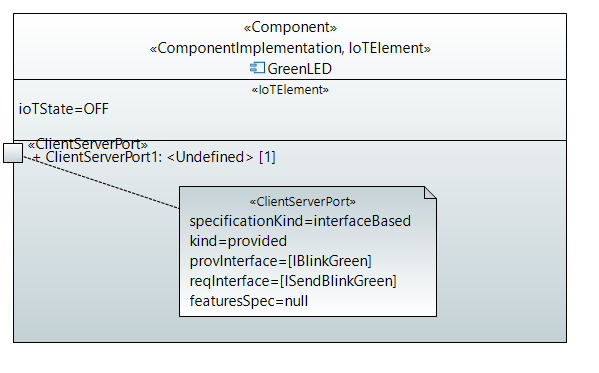}
\caption{LED internal structure}
\label{fig:LEDcsd}
\vspace{-.6cm}
\end{figure}

The Red/Green LEDs have only one unique port to interact with the \textit{Board}. The ports provide the blinking interface but they also require the \textit{“ISendBlinkRed \& ISendBlinkGreen”} from the board respectively. The following figure depicts the internal structure of an LED showcasing the internal structure of ports. Figure \ref{fig:LEDcsd} shows the internal structure of an LED component.

\subsubsection{Ultrasonic sensor}
\begin{figure}[h]
\centering
\includegraphics[width=0.9\linewidth]{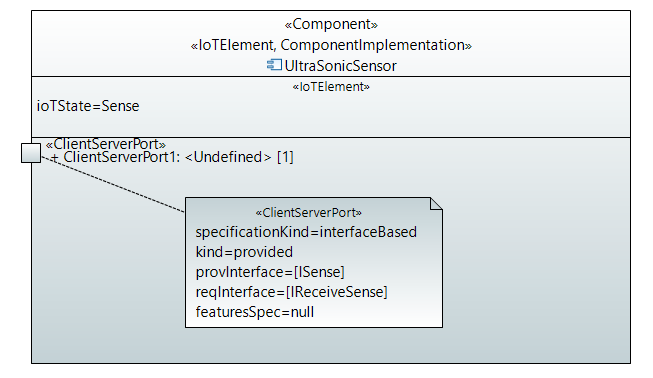}
\caption{Ultrasonic sensor internal structure}
\label{fig:sensorCSD}
\vspace{-.3cm}
\end{figure}

The same as the LEDs, the sensor port also will provide the \textit{“ISense”} interface and require the \textit{“IReceiveSense”} from the board. The figure \ref{fig:sensorCSD} diagram shows how the ultrasonic sensor internal structure.

\subsubsection{A node}
\begin{figure}[h]
\centering
\includegraphics[width=\linewidth]{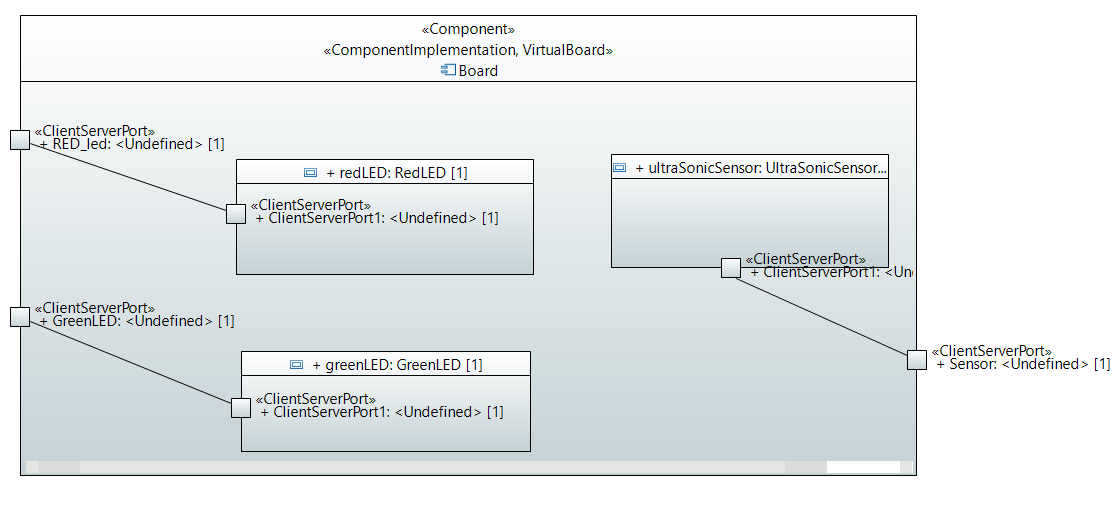}
\caption{Node internal structure}
\label{fig:nodeCSD}
\vspace{-.3cm}
\end{figure}

Modeling the internal structure of a node is done on top of a board and reuse the \textit{IoTElements} that have already been modeled before. Figure \ref{fig:nodeCSD} shows the internal structure of a board.

\textit{\textbf{Note:}} The instantiating other \textit{IoTElements} inside the board component is done by dragging them from the model explorer to the board component in the editor. After dragging them, it is more likely that the ports won’t show. The easiest way to show the component’s content is to: \textit{Select the component → Right Click→ Filter→ Show/Hide content}. From there, select the border ports you want to show. 

\subsection{Modeling components behaviors}

Modeling of component’s behaviours is done using UML state machines, events, guards, and actions. In our approach, where it is necessary, each component should have its own internal state machine describing its behaviour. To start with, each component always has an initial declared state while each \textit{state} should have at least one event of type \textit{“IoTEvent”} to be executed either \textit{“OnEntry, OnExit or continuously”}. In the following part, we will go over all the components' behaviour specifications. But before continuing, let us first familiarize ourselves with different behavioral terminologies proposed by CHESSIoT.

\textbf{Payload}: This is a standalone and straightforward object to carry information to be passed between components. In CHESSIoT, the payload can have zero or many primitive or derived properties to be defined in a message. For instance, suppose a component message to be communicated among components contains a string value, an integer, or even an instance of another payload. 

\textbf{IoTState}: This serves to keep the component state from its initial participation until its disposal in the system. \textit{IoTState} extends actual UML states but in addition to that, \textit{IoTState} carries information related to what events need to be taken care of at a certain point in time. For instance, \textit{OnEntry} or \textit{OnExit} events are triggered when entering or exiting a state. An IoT state can also trigger an internal event, which corresponds to an internal action to be taken. 

\textbf{IoTEvent}: Events in CHESSIoT are triggered in a different manner depending on the state of the component. An \textit{IoTvent} can be incoming, outgoing, or generic, which means it can come from the inside-out, from the outside, or internal. A \textit{GenericEvent} is an event that gets triggered internally to the component, for example, changing variable value.

\textbf{IoTAction}: IoTAction(s) can be of different types depending on the kind of action to be performed. For instance, the \textit{SendPayload} action is referred to when an \textit{OutgoingEvent} is triggered to send the payload through a specified port while \textit{ReceivePayload} is used on \textit{IncomingEvent}s to receive messages from another components. A \textit{GenericAction} does not require to access the component's ports, for example, changing the component's property value.

\textbf{State Transition}: In CHESSIoT, state transitions enable transiting from the source state to a target state, abiding the trigger from the guard value. Guard expressions are boolean expressions defined based on state values. They serve to initiate a state transition by checking whether the \textit{OnExit} event has been performed correctly.

Two stages are involved when modeling the behaviour of the component. First, the definition of payload, events, and actions that a component will use. This is done by using the component inner class diagram. The second step is to create an internal state machine diagram which will then make use of events and actions defined above. The next part details the behavior specification of the node components using a state machine. 

\subsubsection{LED's state machine, event and actions}
\begin{figure}[h]
\centering
\includegraphics[width=\linewidth]{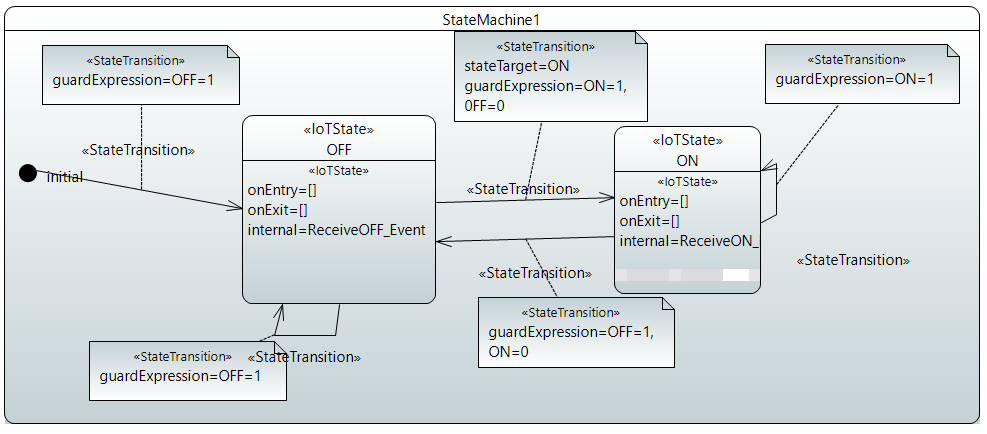}
\caption{LED state machine}
\label{fig:LEDsm}
\end{figure}

As shown in the figure \ref{fig:LEDsm}, an LED has only two continuous states, \textit{“ON”} and \textit{“OFF”}. To trigger transition from one state to another or to self, a guard condition must be fulfilled.

\begin{figure}[h]
\centering
\includegraphics[width=\linewidth]{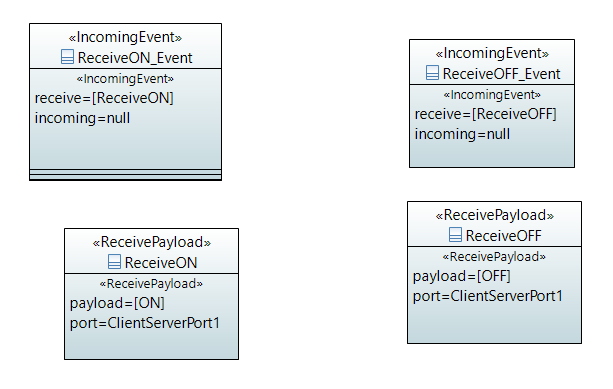}
\caption{LED event and actions}
\label{fig:LEDEvents}
\vspace{-.6cm}
\end{figure}

The only \textit{IoTEvent} an LED executes is an incoming event to receive the payload from the board containing the state of either being low or high (OFF/ON). Figure \ref{fig:LEDEvents} shows the LED's events and action definition.

\subsubsection{Ultrasonic Sensor state machine, event, and actions}

\begin{figure}[h]
\centering
\includegraphics[width=\linewidth]{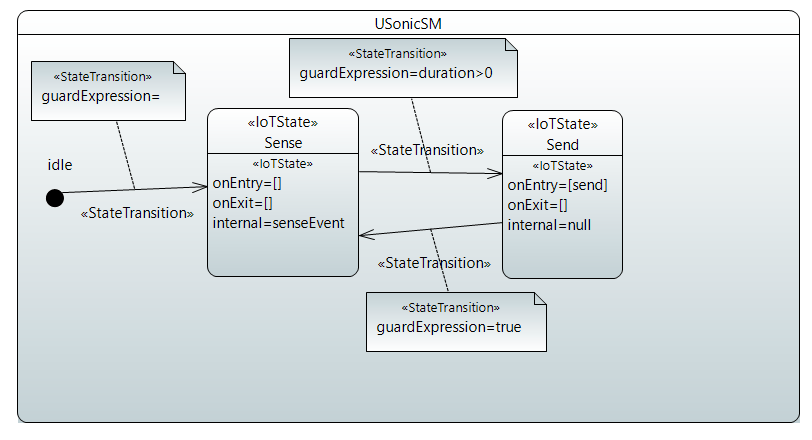}
\caption{Sensor state machine}
\label{fig:sensorSM}
\vspace{-.3cm}
\end{figure}

\begin{figure}[h]
\centering
\includegraphics[width=\linewidth]{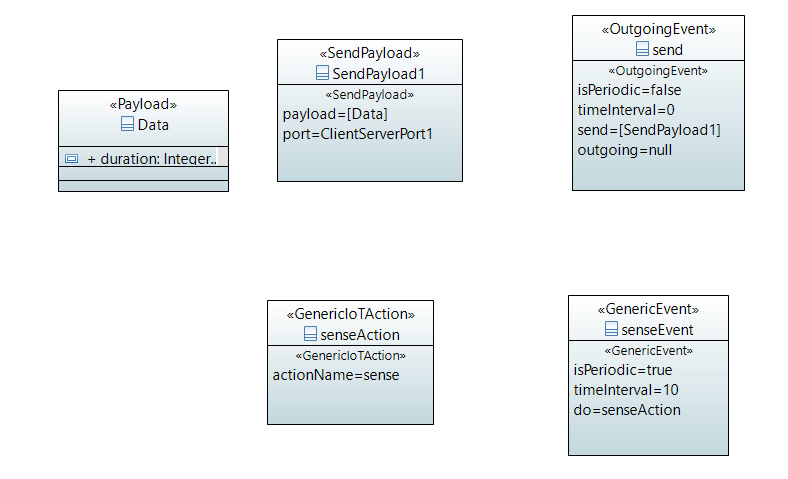}
\caption{Sensor event and actions}
\label{fig:SensorEvents}
\end{figure}

From the figure \ref{fig:sensorSM}, the Ultrasonic sensor contains 2 continuous states \textit{“SENSE”} for sensing and \textit{“SEND”} for sending the payload to the board. So it will use the \textit{"internalEvent"} event type and \textit{"genericAction"} action type for sensing and later use an \textit{"outgoinEvent"} and \textit{“SendPayload”} action for sending the value to the board. The state transitions are triggered by the fulfilling their corresponding guards.
Figure \ref{fig:SensorEvents} shows the sensor's events and action definition.

\subsubsection{Node state machine, event, and actions}
\begin{figure}[h]
\centering
\includegraphics[width=\linewidth]{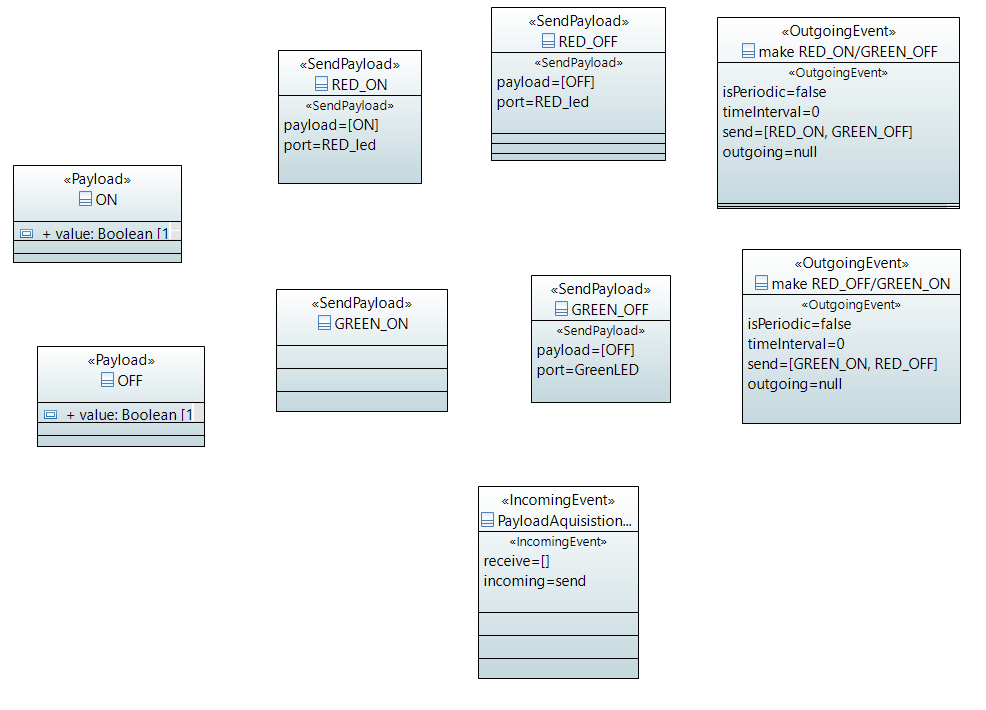}
\caption{Node event and actions}
\label{fig:nodeEvents}
\end{figure}

The figures \ref{fig:nodeEvents} depict the node's behavioural specification (which in turn can be simply referred to as a board). The three main states of a node are \textit{“ACQUISITION”}, \textit{”RED\_ON/GREEN\_OFF}" and \textit{“RED\_OFF/GREEN\_ON"}. Figure \ref{fig:nodeEvents} shows the board's events and action definition. The main triggering condition for the state change is the duration value from the payload. When the time taken by the echo wave to come is greater or equal to 300ms then the space is vacant, while when the echo-wave arrival duration is less than 300ms that means there is a blocking object then the slot is occupied.  The 300ms value is arbitrary, it may be adjusted depending on the position of the sensor.

\begin{figure}[h]
\centering
\includegraphics[width=\linewidth]{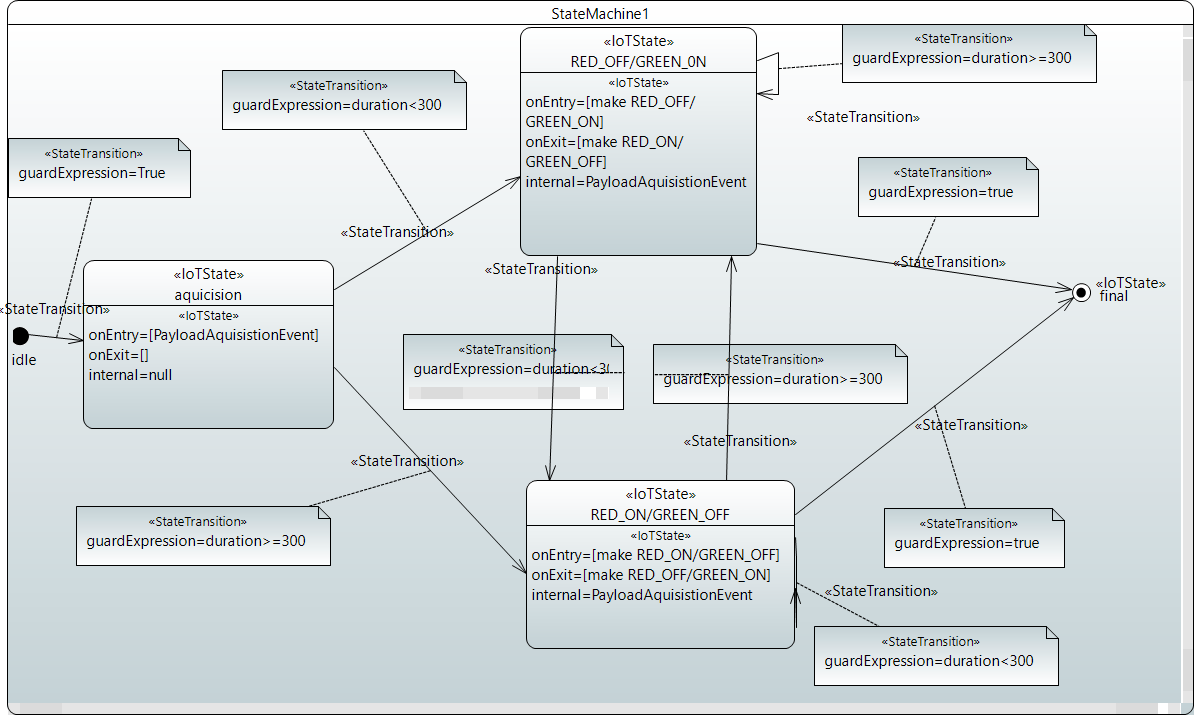}
\caption{Node state machine}
\label{fig:nodeSM}
\vspace{-.3cm}
\end{figure}

The purpose of modeling the behavioral aspect of a system in CHESSIoT is to late generated the platform-specific code. This will be achieved by employing the ThingML platform. The ThingML language combines well-proven software modeling constructs aligned with UML (state-charts and components) and an imperative platform-independent action language to construct the intended IoT applications \citep{LCE4IoT}. ThingML code generator targets many popular programming languages such as C/C++, Java, and Javascript, and about ten different target platforms (ranging from tiny 8bit microcontrollers to servers) and ten different communication protocols \citep{ThingMLcore}.

\section{Conclusion}

Event-based modeling in IoT systems helps to discover the system's hidden behaviors before the development of the system by itself. In this paper, we showcased the capability of the CHESSIoT extension on modeling the behavioral aspects of an IoT system. In the future, CHESSIoT will support the generation of platform-specific code by employing the ThingML tool. Different IoT-specific real-time analyses will be supported too.

\section*{Acknowledgments}
This work has received funding from the Lowcomote project under European Union’s Horizon 2020 research and innovation program under the Marie Skłodowska-Curie grant agreement n\si{\degree} 813884. 

\bibliography{sample}

\end{document}